\documentclass[usenatbib]{basi}
\usepackage[british]{babel}
\usepackage[varg]{txfonts}
\usepackage{rotating}
\usepackage{dcolumn}
\begin{document}
\title[Radio Transients]{Radio Transients: An antediluvian review}

\author[R.P. Fender and M.E. Bell]%
       {R.~P.~Fender\thanks{email: \texttt{r.fender@soton.ac.uk}},
       and M.~E.~Bell\thanks{email: \texttt{meb1w07@soton.ac.uk}},\\
       School of Physics \& Astronomy, University of Southampton\\}

\pubyear{2011}
\volume{00}
\pagerange{\pageref{firstpage}--\pageref{lastpage}}

\date{Received \today}

\maketitle
\label{firstpage}

\begin{abstract}
We are at the dawn of a new golden age for radio
astronomy, with a new generation of facilities under construction and
the global community focussed on the Square Kilometre Array as its
goal for the next decade. These new facilities offer orders of
magnitude improvements in survey speed compared to existing radio
telescopes and arrays. Futhermore, the study of transient and variable
radio sources, and what they can tell us about the extremes of
astrophysics as well as the state of the diffuse intervening media,
have been embraced as key science projects for these new facilities.
In this paper we review the studies of the populations of
radio transients made to date, largely based upon archival surveys.
Many of these radio transients and variables have been found in the
image plane, and their astrophysical origin remains unclear.  We 
take this population and combine it with sensitivity estimates for the
next generation arrays to demonstrate that in the coming decade we may
find ourselves detecting $10^5$ image plane radio transients per year,
providing a vast and rich field of research and an almost limitless
set of targets for multiwavelength follow up.
\end{abstract}



\section{Radio transients: the potential}

The Universe is a violent and dynamic environment, in which the
explosions of massive stars can outshine an entire galaxy,
supermassive black holes swallow stars whole, merging neutron stars
cause ripples in the fabric of spacetime and bursts of ultra-high
energy radiation which can be detected at vast distances, and
particles are accelerated to energies far surpassing anything possible
in laboratories on the Earth. The extremes of physics -- density,
temperature, pressure, velocity, gravitational and magnetic fields --
experienced in these environments provide a unique glimpse at the laws
of physics operating in extraordinary regimes.  Such `extreme
astrophysics', is a high priority for global research in the 21st
century\footnote{See e.g. the ASTRONET report {\em A Science Vision
    for European Astronomy} available at {\bf www.astronet-eu.org}}.
In the second half of the 20th century, astronomy moved beyonds its
optical origins into the radio, infrared, ultraviolet, X-ray and
gamma-ray bands. The most extreme environments in the Universe betray
themselves by their copious high-energy emission, in the X-ray and
gamma-ray bands, and also by their radio emission. Strong magnetic
fields and shock waves as matter collides in the ambient medium,
result in radio bursts from electrons (and positrons) spiralling
around, and sometimes being channelled along, these magnetic field
lines. Wide-field orbiting X- and gamma-ray observatories, with
all-sky monitors, have revealed our Galaxy to be a hotbed of explosive
and relativistic effects, which go largely unnoticed by traditional
optical telescopes both because they are missed in the narrow fields
of view, and are obscured by interstellar dust.  By far the best way
to observe and understand the violent Universe from the ground is via
the radio emission which is ubiquitously associated with these violent
events. The emission arises as the shocked particles glow with
synchrotron radiation, streaming along and around magnetic field
lines, or are induced to move and emit coherently, producing extremely
bright bursts. However, in the past radio astronomy, like optical
astronomy, has suffered from the narrow fields of view which are a
consequence of large dishes, and cannot survey the sky sufficiently
rapidly to detect rare and rapid events, which may be the most
significant astrophysically.  All of this is about to change, with a
new generation of radio telescopes, taking advantage of new technology
and the vast advances made in data transport and computer processing,
connecting large numbers of small detectors, able to deliver enormous
fields of view. Finally the age of the radio all-sky monitors is
here. What is more, these radio all-sky monitors will have several
advantages over the traditional X-ray and gamma-ray monitors: they
will be able to localise events immediately with arcsec precision, and
will be able to go significantly deeper / further.

As we shall demonstrate in this review, our current understanding of
the population of radio transients implies that we should detect
thousands, or even millions, of such events with `next generation'
radio facilities such as The Low-Frequency Array (LOFAR), The South
African Karoo Array Telescope (MeerKAT) and The Australian SKA
Pathfinder (ASKAP), and ultimately the Square Kilometre Array (SKA).
As well as drawing our attention to the extremes of astrophysics,
these sources are potentially observable to vast distances (possibly
as far as the Epoch of Reionisation at redshift $z > 6$) and will turn
out to be amongst our most valuable probes of the intergalactic and
intercluster medium on large scales. They may furthermore, in some
cases, turn out to be the electromagnetic signatures of events which
produce detectable gravitational waves, which would be a breakthrough
connection.

\section{Transient and variable source populations}

In the following subsections, we shall briefly discuss the diverse
populations of transient and variable radio sources, which we separate
into incoherent and coherent processes. These two source populations
also divide, roughly, in the technique which is optimised for their
discovery: incoherent synchrotron transients will be mostly discovered
and studied in the image plane, and coherent sources in time series
analyses using in large part the methods traditionally used for radio
pulsars. There is, it should be noted, some considerable overlap
between the two, in particular as many coherent sources are likely to
be strong, long-lasting and/or dispersed enough to be detectable in
images.  The incoherent sources are also referred to, somewhat
disparagingly, as `slow' transients, and the coherent sources as `fast'
(Cordes 2007).

\subsection{Synchrotron sources}

All astrophysical events which are associated with the explosive
injection of energy into an ambient medium result in synchrotron
emission. The explosive event creates shocks which accelerate
particles to extremely high energies (sometimes above 10$^{19}$ eV),
as well as compressing and amplifying the ambient magnetic field. The
accelerated ultrarelativistic particles spiral around the magnetic
field lines, emitting polarised synchrotron radiation as they lose
energy (see e.g. Longair 1994). A wide variety of astrophysical phenomena are associated with
synchrotron emission, including a menagerie of explosive events such
as:

\begin{itemize}
\item{
supernovae resulting from the collapse of the iron core of a massive
star, or the conflagration of an entire white dwarf as it reaches the
Chandresekar limit (Weiler et al. 2002);}
\item{giant outburst from magnetars, powered by the rearrangement of the
magnetic field of super highly-magnetised neutron stars (Gaensler et al. 2005);}
\item{classical nova explosions, caused by a wave of nuclear fusion
  across the entire surface of a white dwarf (e.g. Pavelin et al. 1993);}
\item{relativistic jets (Hughes 1991), powerful outflows of kinetic
  energy and matter launched from the cores of accreting black hole
  and neutron star systems (e.g. Fender 2006); this phenomenon, which
  is vital to black hole accretion and feedback, is as yet poorly
  understood and yet seen across a broad range of objects from
  stellar-mass to supermassive black holes, and even gamma-ray bursts
  (e.g. Frail et al. 2003), the most powerful explosions in the
  Universe.}
\end{itemize}

\begin{figure}
\centerline{\includegraphics[width=9cm, angle=-90]{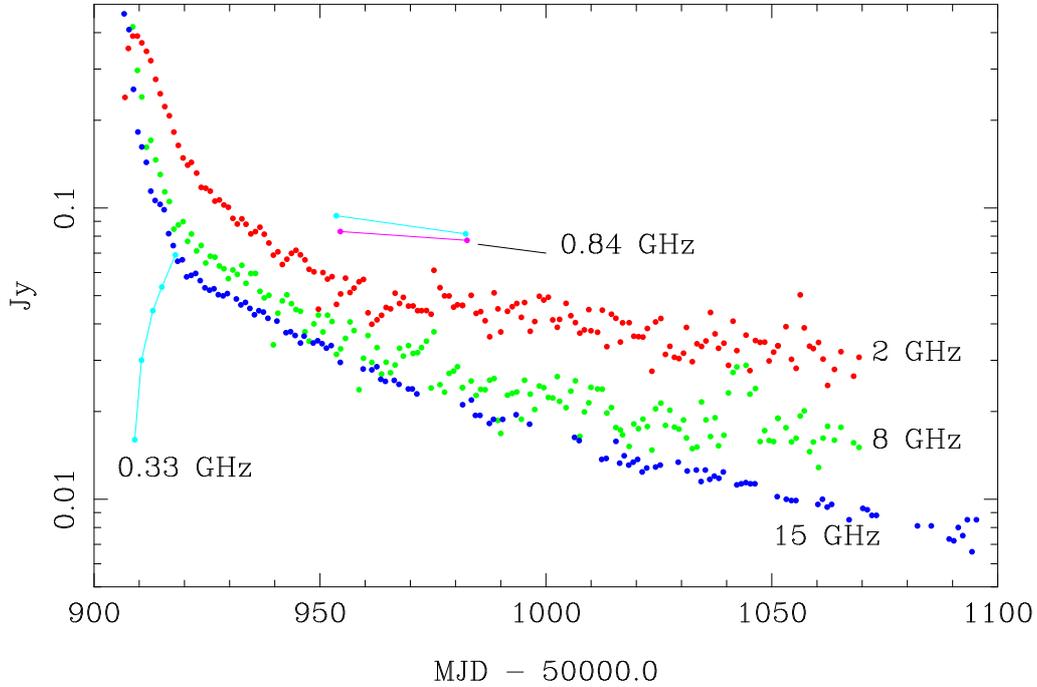}}
\caption{An outburst of the binary transient CI Cam observed at radio
  wavelengths. This is incoherent synchrotron emission from electrons
  shock-accelerated by the ejection of matter from the binary system
  (in a jet and/or wind). The initial rise phase corresponds to a
  combination of the particle acceleration timescale (which may
  dominate at high frequencies) and the evolution of initially
  self-absorbed emission to optically thin (which clearly dominates
  below 1 GHz).}
\label{cicam}
\end{figure}

Fig.~\ref{cicam} illustrates an incoherent synchrotron outburst from a
galactic binary source. The lower ($< 1$ GHz) frequencies peak lower and
later than the higher (GHz) frequencies, due to evolving source size
and optical depth. Note that the duration of such events depends upon
the rate at which their volume changes significantly. For similar
(relativistic or mildly relativistic) expansion velocities, this means
that less luminous events (e.g. flares from cataclysmic variables or
X-ray binaries (such as that in Fig.~\ref{cicam}) will evolve much more
rapidly than more luminous events such as SNe, GRB afterglows and AGN
outbursts (which can of course be seen to much larger distances).  It
is also worth noting, although it is obvious, that the decay phase
lasts much longer than the rise phase (even at MHz frequencies), and
so the radio sky may turn out to be dominated by faint sources which
are slowly fading (e.g. thousands of fading GRB afterglows across the
sky).

\subsection{Coherent bursts}

Short-timescale (often less than one second, sometimes unresolved on
millisecond timescales) `coherent' transients are also often
associated with the some of the highest energy density events in the
Universe. However, unlike synchrotron radiation, which in a steady
state is limited to a brightness temperature of $\sim 10^{12}$K, for
the most extreme coherent events the brightness temperature is in
excess of 10$^{35}$ K. Pulsar emission is perhaps the most famous form
of coherent radio emission, and has unequivocally demonstrated its
astrophysical importance by demonstrating the existence of neutron
stars (Hewish et al. 1968, leading to the Nobel Prize 1974) and
providing exquisitely precise tests of general relativity (Hulse and
Taylor 1975, leading to the Nobel Prize, 1993). The same coherent
emission is also associated with other relatives of the standard radio
pulsars, such as the RRATs and intermittent pulsars (McLaughlin et al. 2006).

\begin{figure}
\centerline{\includegraphics[width=13cm]{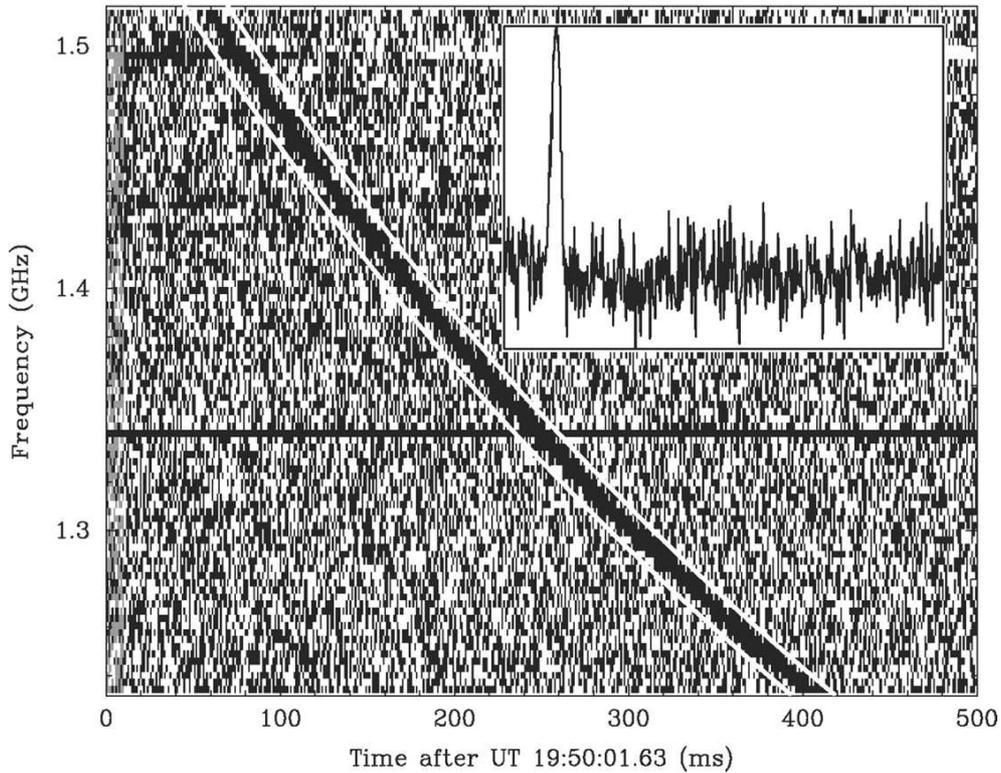}}
\caption{The observation of a possible coherent radio burst at a
  distance of $\sim$500 Mpc (Lorimer et al. 2007), showing the
  quadratic dispersion signature expected for an event of
  astrophysical origin (but see Burke-Spoloar et al. 2011 and Keane et
  al. 2011). Such bursts could potentially be seen as far back as the
  first population of luminous objects in the epoch of reionisation,
  and could be unparalleled tools for the study of the intergalactic
  and intercluster medium on the largest scales. Reprinted with
  permission from AAS. }
\label{lorimer}
\end{figure}

In 2007, Lorimer et al. (2007) announced the possible discovery of a
short duration, dispersed burst from an estimated distance of 500 Mpc
(Fig.~\ref{lorimer}).  They furthermore estimated that there should be
hundreds of such bursts per day across the sky. Although there remains
some doubt as to the astrophysical origin of the burst (Burke-Spoloar
et al. 2011, but see Keane et al. 2011 for other possible
events), such bursts would open up an entirely new field of
astrophysics, acting as probes of the intergalactic medium (IGM) by
measuring pulse dispersion and broadening, providing direct measures
of the IGM magnetic field and turbulence.

For the rest of this review we shall focus on current best estimates
of the radio transients population, and how it maps onto the
sensitivities of the new generation of radio facilities. Nearly all of
this work is done in the image plane and corresponds to `slow'
transients. For much more on fast transients we refer the reader to
e.g. Cordes, Lazio \& McLaughlin (2004), Cordes (2007), Hessels et
al. (2009), Stappers et al. (2011), Keane \& McLaughlin (2011).

\section{The current situation}

Despite the scientific potential, the transient and time variable
radio sky is a relatively unexplored region of parameter space. The
restricted survey speeds of older generation radio facilities has
meant that detections of radio transients lag far behind that which is
possible with, for example, X- and gamma-ray instruments.  Four
dominant methods have so far been used to detect radio transients (i)
dedicated surveys (ii) multi-wavelength triggered detections (iii)
archival studies and (iv) serendipitous and commensal
detections. Commensal refers to the process of searching numerous wide
field or deep observations, which have been obtained to address a
number of scientific goals. Commensal observations can provide many
thousand of hours of data without interrupting an over subscribed
telescope schedule (discussed and quantified further later).  Some
radio transients have been detected serendipitously over the years,
but they so far remain relatively limited in number (for examples see
Davies et al. 1976; Zhao et al. 1992; van den Oord \& de Bruyn 1994;
Bower et al. 2003 and Lenc et al. 2008).

\subsection{Dedicated surveys}

The first well studied variables in the mid to late 20$^{th}$ century
were typically bright (often 3C) extragalactic radio sources (for
examples see Hunstead 1972; Condon \& Backer 1975; Ryle et al. 1978;
Condon et al. 1979; Dennison et al. 1981; Taylor \& Gregory 1983;
Aller et al. 1985; Simonetti \& Cordes 1990; Riley et al. 1993; Lovell
et al. 2008). The focus at the time was primarily placed on studying
the intrinsic properties (including scintillation) of these known
sources, rather than exploring parameter space. Some of the papers
above often refer to the fractional change (including alternative
metrics of variability) in the flux of a source on a specific
timescale, rather than the number of sources (per deg$^{2}$) whose
flux changed by a given amount on a specific timescale. The latter is
more useful for understanding the global population i.e. the number of
sources which are extremely variable in a given flux and time interval, per
solid angle on the sky.

As well as distant extragalactic radio sources, more recently nearby
galaxies have been common targets for transient surveys. For example,
M82 hosts a large number of supernovae remants due to a prolific
star formation rate; a number of unidentified radio transients have
also been detected (Kronberg \& Sramek 1985; Muxlow et al. 1994;
Brunthaler et al. 2009; Joseph et al. 2011).  The Galactic centre 
has also been the area for some intense observing campaigns.  These
surveys have so far detected a number of radio transients (Davies et
al. 1976; Hyman et al. 2002; Bower et al. 2005; Hyman et al. 2006;
Hyman et al. 2007; Hyman et al. 2009). 

In recent years the focus has also shifted towards performing (truly)
dedicated blind surveys of parameter space to search for radio
transients (i.e. not necessarily targeting hotbeds of nearby transient
activity). These surveys have also focused on parameterising the rate
of events on the sky (a discussion of which will follow later). For
example, nine bursts -- named the WJN transients -- in excess of 1 Jy
have been reported from  drift scan observations with the Waseda
Nasu Pulsar Observatory at 1.4 GHz, which are summarised in Matsumura
et al. (2009) (also see Kuniyoshi et al. 2007; Niinuma et al. 2007;
Kida et al. 2008; Niinuma et al. 2009 for further details). These are
some of the brightest slow transients (lasting minutes to days) which
have been reported in the literature: they so far remain unexplained.
Croft et al. (2010) published results from the dedicated Allen
Telescope Array Twenty Centimetre Survey (ATATS): no unique transients
were detected and an upper limit on the snapshot rate of events was
given. A follow-up paper by Croft et al. (2011) further constrained the
rates of radio transients; both papers also present a comprehensive
study of the variability of persistent sources within the
observations. Subsequently the Pi GHz Sky Survey (PiGSS) surveyed the
sky with the ATA at 3.1 GHz, providing the deepest static source
catalogue to date above 1.4 GHz (Bower et al. 2010).  No unique radio
transient sources were reported in this survey and an upper limit on
snapshot rate was placed (also see Bower et al. 2011). In the low
frequency regime Lazio et al. (2010) have used the Long Wavelength
Demonstrator Array (LWDA) at 73.8 MHz to perform an all-sky survey for
bright ($>500$ Jy) moderately fast ($\sim$ 5 minutes) transients. No
astronomical sources are detected which places a stringent limit on
the rate of extremely bright ($\geq 500$ Jy) low frequency events.

In conjunction with searching for unique (previously unknown)
transient radio sources, searching for highly variable faint radio
sources can also be a useful diagnostic in examining the dynamic radio
sky. For example, Carilli et al. (2003) found a number of highly
variable ($\Delta S \geq\pm 50$\%) radio sources in a small number of
repeated observations of the Lockman Hole at 1.4 GHz. Frail et
al. (2003) found four highly variable radio transient sources from
follow-up observations of GRBs at 5 and 8.5 GHz. Reporting 39 variable
radio sources, Becker et al. (2010) characterised the surface density
of variables in the direction of the Galactic plane at 4.8 GHz.  The
majority of the variable sources presented in Becker et al. (2010) had
no known multi-wavelength counterparts. Ofek et al. (2011) also perform
a blind search for transients at low Galactic latitudes. This study is
one of the first to characterise the variability of sources on
pre-described cadences (days, months and years).

\subsection{Multi-wavelength triggered detections}

Since the development of high energy observatories detections of radio
transients have often relied on multi-wavelength triggered
observations.  These have produced radio counterparts to gamma-ray
burst (GRB) afterglows, Soft Gamma-Ray Repeaters (SGRs) and extremely
rich datasets of black hole X-ray binary outbursts (for
examples see Frail et al. 1997; Eck, Cowan \& Branch 2002; Gaensler et
al. 2005; Fender, Homan \& Belloni 2009).  This method relies on
having a detectable high frequency counterpart, which may be absent
(or difficult to detect) for sources such as X-ray dim isolated
neutron stars (XDINs; see Ofek et al. 2009 for discussion) and orphan
gamma-ray burst afterglows (Frail et al., 1997).

\begin{figure}
\centerline{\includegraphics[width=13cm]{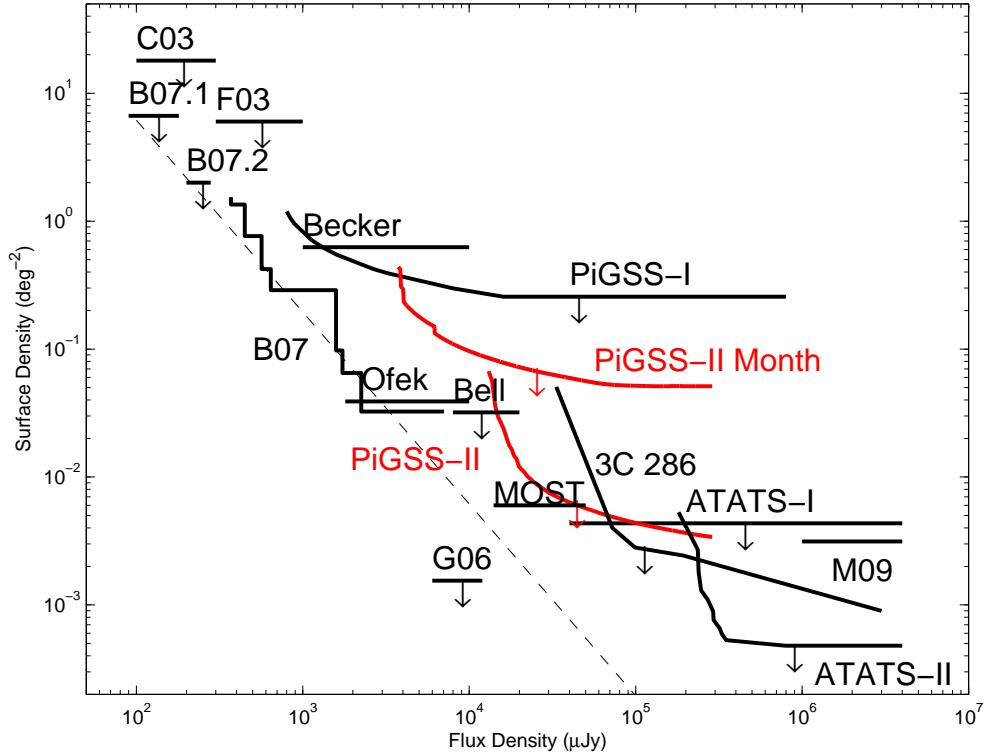}}
\caption{The current state of the art $\log N$ -- $\log S$
  distribution for GHz-frequency radio transients with variability
  timescales of weeks (from Bower et al. 2011). While most surveys
  have resulted in limits on source populations, the definitive study
  remains that of Bower et al. (2007), for which the slope of the
  population (indicated by `B07' in the figure) is consistent with a
  homogenous source population in Euclidean space. Fig.~\ref{SKA_plot}
  extends this figure to include the MHz results of Lazio et
  al. (2010) and demonstrates how, based upon the Bower population,
  the next-generation facilities should discover very large numbers of such
  sources. Reproduced by permission of the AAS. }
\label{bower}
\end{figure}

\subsection{Archival studies}

Radio telescope archives potentially contain many hours of data which
currently remains to be searched for radio transients. An archival
study comparing the NVSS (NRAO VLA Sky Survey; Condon et al. 1998) and
FIRST (Faint Images of the Radio Sky at Twenty-cm; Becker et al. 1995;
White et al. 1997) catalogues was conducted by Levinson et al. (2002)
to constrain the rates of orphaned GRB afterglows. A follow-up study
was conducted by Gal-Yam et al. (2006) and a number of radio transient
sources were identified.  Thyagarajan et al. (2011) have recently
performed the most thorough analysis of the FIRST survey to date,
reporting 1627 transient and variable sources down to mJy levels, over
timescales of minutes to years. Thyagarajan et al. (2011) report that
-- so far -- 877 of the detected variable sources lack optical
counterparts.  See also de Vries et al. (2004) and Ofek \& Frail
(2011) for further analysis and discussion of the FIRST and NVSS
catalogues.

In what remains the benchmark study, Bower et al. (2007) analysed 944
epochs of archival VLA data at 4.8 and 8.4 GHz spanning a period of 22
years. In this survey ten radio transients were reported, with the
host galaxies possibly identified for four out of the ten sources, and
the hosts and progenitors of the other six unknown. Croft et al. (2011)
perform X-ray follow-up of the transient sources reported in Bower et
al. (2007), from which eight sources still remain completely
undetected. Bannister et al. (2011) published results from a search for
transient and variable sources in the Molonglo Observatory Synthesis
Telescope (MOST) archive at 843 MHz: 15 transient and 53 highly
variable sources were detected over a 22 year period. Bannister et
al. (2011) use these detections to place limits on the rates of
transient and variable sources. Bower \& Saul (2010) and Bell at
al. (2011) have published further archival work examining observations
of the VLA calibrator fields. The calibrators are the most heavily
observed VLA fields and are rarely imaged for scientific
purposes. Both these studies placed constraints on the rates of
transients at GHz frequencies and each study analysed thousands of
images respectively. Also, Gaensler \& Hunstead (2000) examined archival
observations of the Molonglo calibrators reporting that 18 out of 55 were
variable on timescales one to ten years.

The NVSS, FIRST (and other legacy surveys) are extremely important for
transient studies because they offer a comparative catalogue to assess
the flux, timescale and robustness of new transient and variable
sources. Two new all-sky legacy surveys are planned (EMU --
Evolutionary Map of the Universe and WODAN -- Westerbork Observations
of the Deep APERTIF Northern-Sky) which will produce both Northern and
Southern hemisphere catalogues down to micro-Janksy sensitivities:
these will provide the bedrock for new transient and variability
studies. These new surveys can also be compared with older legacy
surveys such as NVSS and SUMSS (Sydney University Molonglo Sky Survey;
Bock et al. 1999) to search for further transient sources.

Archival studies are hampered by the inability to perform
instantaneous real time follow-up, which is often needed to correctly
classify a source. So far, a large fraction of the reported transients
appear to lack optical and X-ray counterparts (Bower et al. 2007;
Matsumura et al. 2009; Thyagarajan et al. 2011), either the sample of
sources so far are indeed optically and X-ray dim (Ofek et al. 2010),
or this is selection effect related to the time between follow-up.
The current state-of-the-art $\log N$--$\log S$ for GHz transients,
from Bower et al. (2011), is presented in Fig.~\ref{bower}.

\section{The future}

Radio astronomy appears to be at the dawn of a new golden age, with
major facilities under construction and older facilities undergoing
dramatic upgrades, driven in large part by the desire to achieve the
full Square Kilometre Array within 15 years (e.g. Carilli \& Rawlings
2004).

In some cases, these new facilities and upgrades are already here, and
their potential just starting to be explored. In the following section
we discuss how, based upon the existing population studies, we expect
the new radio telescopes to perform as transient detection machines.

\subsection{Estimated transient detection rates}

\begin{figure}
\centerline{\includegraphics[width=14cm]{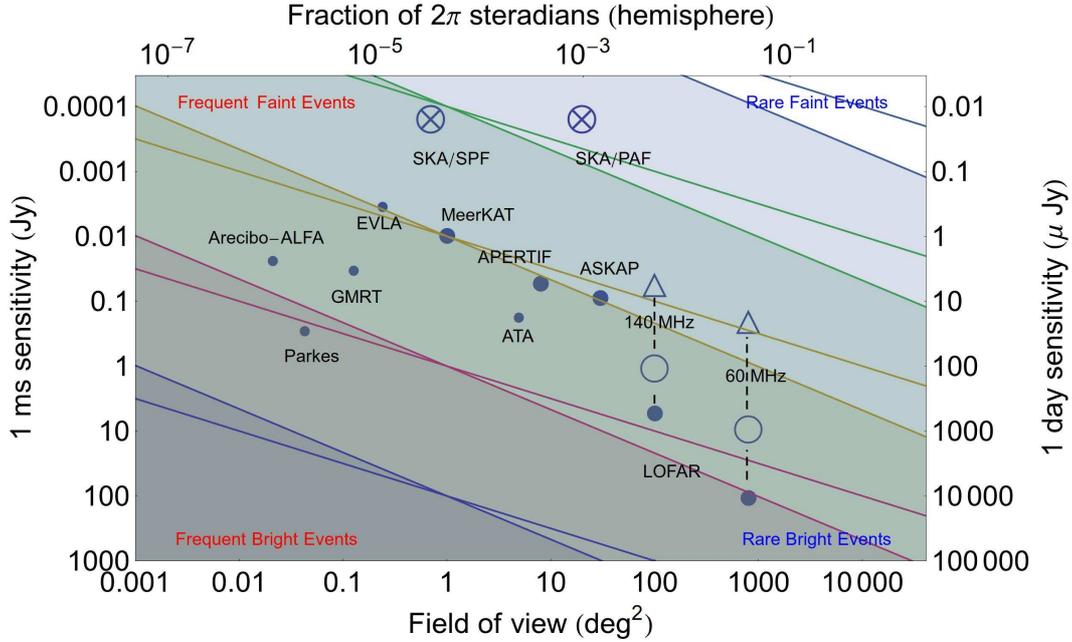}}
\caption{Illustration of the sensitivities and fields of view of some
  of the current and next generation of radio telescopes. For LOFAR
  which operates at much lower frequencies than the other facilities
  (which, for the purposes of this figure, we assume are all operated
  in the 1.4 GHz `L' band), we show both the 140 MHz and 60 MHz bands,
  each with raw sensitivities and two sets of spectral connections
  (connected by the vertical dashed lines).  The open circles
  represent a correction for a spectral index of -0.7, expected for
  optically thin synchrotron sources at low frequencies. The open
  triangle represents a lower limit on corrections for a population of
  coherent sources, which are likely to be at least as steep as
  spectral index -2. These scenarios are also reproduced in the source
  count estimates in Table 1.  Also indicated are lines indicating a
  constant number of detected sources for a homogenous population of
  sources in a Euclidean Universe (FoV $\propto$ sensitivity$^{2/3}$)
  in a single observation, as well as the more traditional survey
  figure of merit, (FoV $\propto$ sensitivity$^{1/2}$) which
  corresponds to the number of square degrees per unit time to a
  uniform depth. These are indicators of likely yields from simple
  pointings and surveys, respectively. It is noteworthy that the new
  generation of GHz-frequency facilities (EVLA, MeerKAT, WSRT+APERTIF
  and ASKAP), despite having different combination of field of view
  and sensitivity, all have comparable survey speeds.  This figure was
  inspired by Fig 1 of Macquart et al. (2010).}
\label{fov}
\end{figure}

For a homogenous set of sources distributed in Euclidean space, we
expect the number of detected sources $N$ per steradian (i.e. the
surface density) to be a function of sensitivity $S$ (smaller $S$ is
more sensitive) such that $N \propto S^{-3/2}$ which is consistent
with the results found in Bower et al. (2007). We can take this
assumption and use it to compare the expected potential for transient
population studies with different telescopes, if we take into account
their fields of view, $\Omega$ i.e.  $N \propto \Omega S^{-3/2}$,
which means that on a plot of sensitivity against field of view,
functions of the form $S \propto \Omega^{2/3}$ correspond to constant
numbers of transients (or, indeed, sources of any kind) found. This
function is useful for comparing estimated transient rates in simple
observations (often performed for other scientific goals).
We present such a plot, with lines drawn indicating constant numbers
of transients, for the next generation radio facilities, together with
a subset of major existing facilities, in Fig.~\ref{fov}.  Of course
such estimates are very crude, taking no account of a heterogenous
source population nor observing strategy.

Note that this measure of sources found differs slightly from the
traditional radio telescope survey figure of merit (FOM), which
corresponds to the number of square degrees per unit time that can be
surveyed to a given flux limit: FOM$ \propto \Omega / \delta t \propto
\Omega S^{-2}$ (since the sensitivity achieved in a given integration
time $\delta t$ is $\propto \delta t^{-1/2}$). This leads to functions
for a constant FOM, on the same figure, of the form: $S \propto
\Omega^{1/2}$. Both functions are indicators of the potential for the
exploration of discovery space with different facilities.

It is clear from Fig.~\ref{fov} that the new / upgraded generation of
GHz facilities (EVLA, MeerKAT, WSRT-APERTIF and ASKAP) could each --
under the naive assumptions made above -- discover comparable number
of transient sources. Viewed this way, the low frequency array LOFAR
does not appear competitive. However, it is important to correct for
the order of magnitude frequency difference between the GHz facilities
and LOFAR. A typical synchrotron source at low radio frequencies, once
optically thin, would have a spectral index $-0.5 \geq \alpha \geq
-1.0$ where $S_{\nu} \propto \nu^{\alpha}$. For $\alpha \sim -0.7$,
LOFAR is comparable to the ATA; for $\alpha \sim -2$, LOFAR is
arguably the best facility (these corrected figures are indicated in
Fig.~\ref{fov} by the open symbols). Although $\alpha \sim -2$ is
steep for a synchrotron source, it is not at all unusual, and arguably
rather shallow, for a coherent source. This, then, demonstrates that
LOFAR, while being a powerful instrument for synchrotron sources, will
be the world-leading facility for coherent sources. This statement
does of course depend rather heavily on the assumption that reasonable
correction can be made for propagation effects (e.g. dispersion) which
are stronger at lower frequencies. It may well transpire that the GHz
and MHz facilities sample populations with quite different fractional
contributions from synchrotron and coherent sources, and a comparison
of the MHz transient population with the Bower et al. GHz distribution
is one of the key early science goals for LOFAR.

\begin{figure}
\centering
\includegraphics[scale=0.52,angle=0]{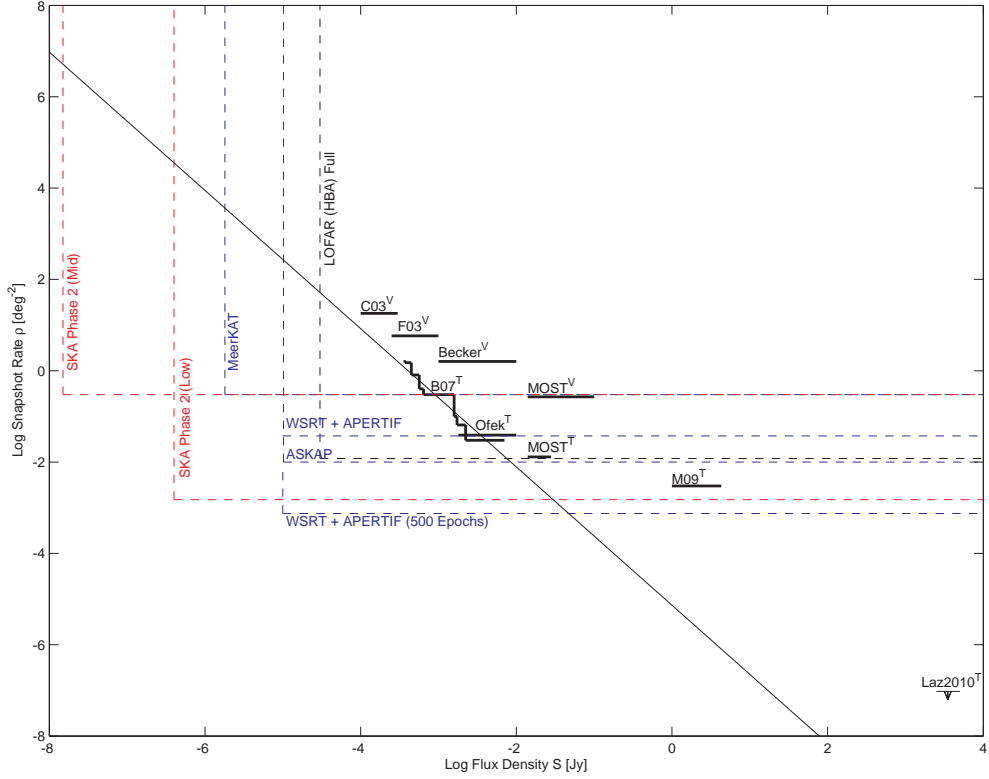}
\caption[Predicted parameter space that the SKA and SKA pathfinders
  may explore]{Log of the Snapshot rate (deg$^{-2}$) against log of
  the flux density (Jy) of detections of transients (labelled `T'),
  detections of variable sources (labelled `V') and upper limits based
  on non-detections (labelled with downward arrows). The dashed lines
  represent the sensitivity (vertical), and 2$\sigma$ upper limits on
  the snapshot rate (horizontal), derived for a number of SKA
  pathfinder instruments, assuming zero detections in 10 epochs of
  data. It is clear that the new facilities will push deep into
  detection territory -- see Table 1 for estimated rates.}
\label{SKA_plot}
\end{figure}

We may look at predictions for the new radio facilities in more
detail.  Fig.~\ref{SKA_plot} shows an adapted version of the flux
density versus snapshot rate plot (see Fig.~\ref{bower}). In this
figure we plot only the blind transient surveys which have detected
transient or variable sources; for clarity we include the Lazio et
al. (2010) upper limit.  These come from the 73.8 MHz Long Wavelength
Demonstrator Array, have the widest field of view of any survey
to date, and are the benchmark against which the first wide-field
LOFAR transient surveys will be measured (as noted before, the MHz and
GHz populations of transients may be rather different). Also shown is
the derived least-squares best fit to the Bower et al. (2007)
detections of 1 week duration (thin black line). 

In Fig.~\ref{SKA_plot} the vertical dashed lines represent the r.m.s.
noise (in a 12 hour observation) for a number of SKA pathfinder
instruments.  The corresponding horizontal lines represent the
2$\sigma$ upper limit on the snapshot rate, derived from 10 epochs of
data with each instrument respectively. For the SKA Phase 2 we adopt
the values quoted in Carilli et al.  (2003); for a frequency of 1.4
GHz, an r.m.s. of 37 nJy, and a Field of View (FoV) of 1 deg$^{2}$ is
realised. For a frequency of 200 MHz (c.f. LOFAR), an r.m.s. of 0.4
$\mu$Jy, and a FoV of $\sim$200 deg$^{2}$ is realised. The Aperture
Tile in Focus (APERTIF) upgrade to the Westerbork Synthesis Radio
Telescope (WSRT) and the ASKAP telescope achieve a similar sensitivity
of $\sim$ 10 $\mu$Jy (at 1.4 GHz) in a 12 hour observation, however
ASKAP will image a FoV of 30 deg$^{2}$, compared to 8 deg$^{2}$ for
APERTIF (see Oosterloo et al. 2010 and Johnston et al. 2007 for
further details).  The MeerKAT telescope is designed for high dynamic
range imaging, and will achieve an r.m.s. of $\sim$1.8 $\mu$Jy in 12
hours, with a FoV of $\sim$1 deg$^{2}$ at 1.4 GHz.

Fig.~\ref{SKA_plot} also shows the potential full array sensitivity
of LOFAR (31 $\mu$Jy), assuming a 12 hour integration at 120 MHz and
10 repeat visits to the same field. A greater FoV can be achieved with
LOFAR by using multiple beams to tile up a larger FoV, however, this
comes with a trade off of sensitivity. Assuming that the transient
population detected by Bower et al. (2007) are in some way
representative of the global population; by either improving the
sensitivity (which pushes the detection limit on Fig.~\ref{SKA_plot}
to the left); \textit{or} by imaging larger FoVs (which pushes the
horizontal snapshot rate limit down), the transient population can be
efficiently explored. Note, in Fig.~\ref{SKA_plot} we do not correct
for any spectral index effects, with respect to the GHz and MHz
surveys and predictions. An order magnitude increase in flux from 1.4
GHz to 150 MHz would be expected for a source with $\alpha=-0.7$.
This in turn would shift the $\log N$ -- $\log S$ of the Bower et al. (2007)
GHz detections to the right. This is probably an over-simplification,
however, it should be considered when plotting fluxes and snapshot
rates derived at two different frequencies.

Taking the Bower et al. (2007) best fit and extrapolating to the SKA
Phase 2 (Mid) sensitivity limit using:

\[
\log \rho = -1.5 \log S_{v}-5.13
\]

\noindent Where $\rho$ is the snapshot rate and $S_{v}$ is the 
detection threshold of the observations. Assuming a 5$\sigma$ 
detection is needed, an huge snapshot rate of 4.4$\times10^{5}$ deg$^{-2}$ 
is predicted for the SKA Phase 2 (Mid). We also summarise the 
predicted snapshot rates for other instruments in Table \ref{rates}. 
The predictions are extremely speculative about the true nature of 
the transient population, however, even if the $\log N$ -- $\log S$ is less 
steep, large numbers of transient detections are still expected.  

The examples above have assumed that only ten repeat visits are
obtained of the same field. Using the proposed 1.4 GHz--band APERTIF
surveys\footnote{http://www.astron.nl/radio-observatory/apertif-eoi-abstracts-and-contact-information}
as an example, we can calculate how many repeat visits we could expect
from a commensal survey. The APERTIF surveys will last for a period of
$\sim$five years and will tackle a number of scientific goals. The
proposed surveys fall into the following three categories:

\begin{itemize}
\item \textbf{Shallow}\\
($1\times 12$ hours) $\times$ 1250 fields.\\
Total: 10,000 deg$^{2}$ revisited once.

\item \textbf{Medium deep}\\
($10\times 12$ hours) $\times$ 125 fields.\\
($12\times 12$ hours) $\times$ 5 fields.\\
Total: 1000 deg$^{2}$ revisited 10 times.

\item \textbf{Deep} \\
($100\times 12$ hours) $\times$ 5 fields.\\
Total: 40 deg$^{2}$ revisited 100 times.\\
($500\times 12$ hours) $\times$ 1 fields.\\
Total: 8 deg$^{2}$ revisited 500 times.
\end{itemize}

The shallow continuum survey represents (near) all sky Northern
hemisphere coverage to a sensitivity of $\sim10\mu$Jy. This survey
will provide one of deepest radio catalogues of the Northern sky at
1.4 GHz and can be compared with existing radio catalogues to identify
transient radio sources. Note, a number of other projects are
proposing all sky surveys, such as pulsars and HI, however these will
be conducted commensally (i.e. Pulsar and HI data can be obtained
simultaneously with continuum data).  The deep surveys offer the
largest number of repeat observations of the same field, potentially
500 revisits. In Fig.~\ref{SKA_plot}, we include the limit on the
snapshot rate that would be found from 500 repeat visits to the same
field (labelled `WSRT + APERTIF (500 Epochs)').  This offers
approximately an order of magnitude improvement on the snapshot rate,
assuming no transient are detected. Different images could also be
averaged together from the deep and medium surveys to improve the
sensitivity.

\begin{table}
\centering
\caption{Predicted rates of radio transients for SKA pathfinder
  instruments.  The calculations are based on the $\log N$ -- $\log S$
  of the Bower et al. (2007) detections at 4.8 GHz. The snapshot rate
  column shows the rate extrapolated to the 3$\sigma$ detection limit
  of each instrument respectively (Note, per deg$^{2}$). The Bower et
  al. (2007) survey had a cadence of $\sim$one week, therefore the
  rate per year column shows the expected rate of transients in 52
  weeks of observing. Note, this is a lower limit because the
  characteristic timescale of the Bower et al. (2007) transients were
  between 20 minutes and one week, we also do not consider any
  recurrence in transient activity. The yield columns give the number
  of transients expected for each instrument over the intrument's FoV.
  The LOFAR rates do not take into account any spectral index
  considerations.  Therefore, assuming a spectral index of
  $\alpha=-0.7$ (synchrotron) and $\alpha=-2$ (coherent) we show the
  corrected rates based on these spectral indexes. The coherent rates
  in particular should be taken with a lot of caution, since they
  depend on both what fraction of the Bower et al. population were
  coherent, and how well we will be able to deal with propagation and
  scattering effects at low radio frequencies.}
\vspace*{0.5cm}
\begin{tabular}{|c|c|c|c|}
\hline 
Instrument & Snapshot rate  & Rate per year & Yield    \\ 
  & (deg$^{-2}$) &  (deg$^{-2}$ yr$^{-1}$) &  (yr$^{-1}$)   \\ 
\hline
SKA Phase 2 (Mid) & 9.7$\times10^{5}$ & 5.0$\times10^{7}$ & 5.0$\times10^{7}$ \\
SKA Phase 2 (Low) & 6.7$\times10^{3}$ & 3.5$\times10^{5}$ & 7.0$\times10^{7}$ \\
MeerKAT           & 6.9$\times10^{2}$ & 3.6$\times10^{4}$ & 3.6$\times10^{4}$ \\
WSRT + APERTIF    & 5.0$\times10^{1}$ & 2.6$\times10^{3}$ & 2.0$\times10^{4}$ \\
ASKAP             & 5.0$\times10^{1}$ & 2.6$\times10^{3}$ & 7.8$\times10^{4}$ \\
LOFAR (HBA) Full  & 9.8               & 5.0$\times10^{2}$ & 1.2$\times10^{4}$\\
\hline 
LOFAR (HBA) Full ($\alpha=-0.7$) & 1.0$\times10^{2}$ & 5.3$\times10^{3}$  & 1.3$\times10^{5}$ \\
LOFAR (HBA) Full ($\alpha=-2$)   & 8.9$\times10^{3}$ & 4.6$\times10^{5}$  & 1.2$\times10^{7}$ \\
SKA Phase 2 (Low)($\alpha=-0.7$) & 5.5$\times10^{4}$ & 2.8$\times10^{6}$  & 5.7$\times10^{8}$ \\
SKA Phase 2 (Low)($\alpha=-2$)   & 2.4$\times10^{6}$ & 1.2$\times10^{8}$  & 2.5$\times10^{10}$ \\
\hline
\end{tabular}
\label{rates}
\end{table}

\begin{figure}
\centering
\includegraphics[scale=0.35,angle=0]{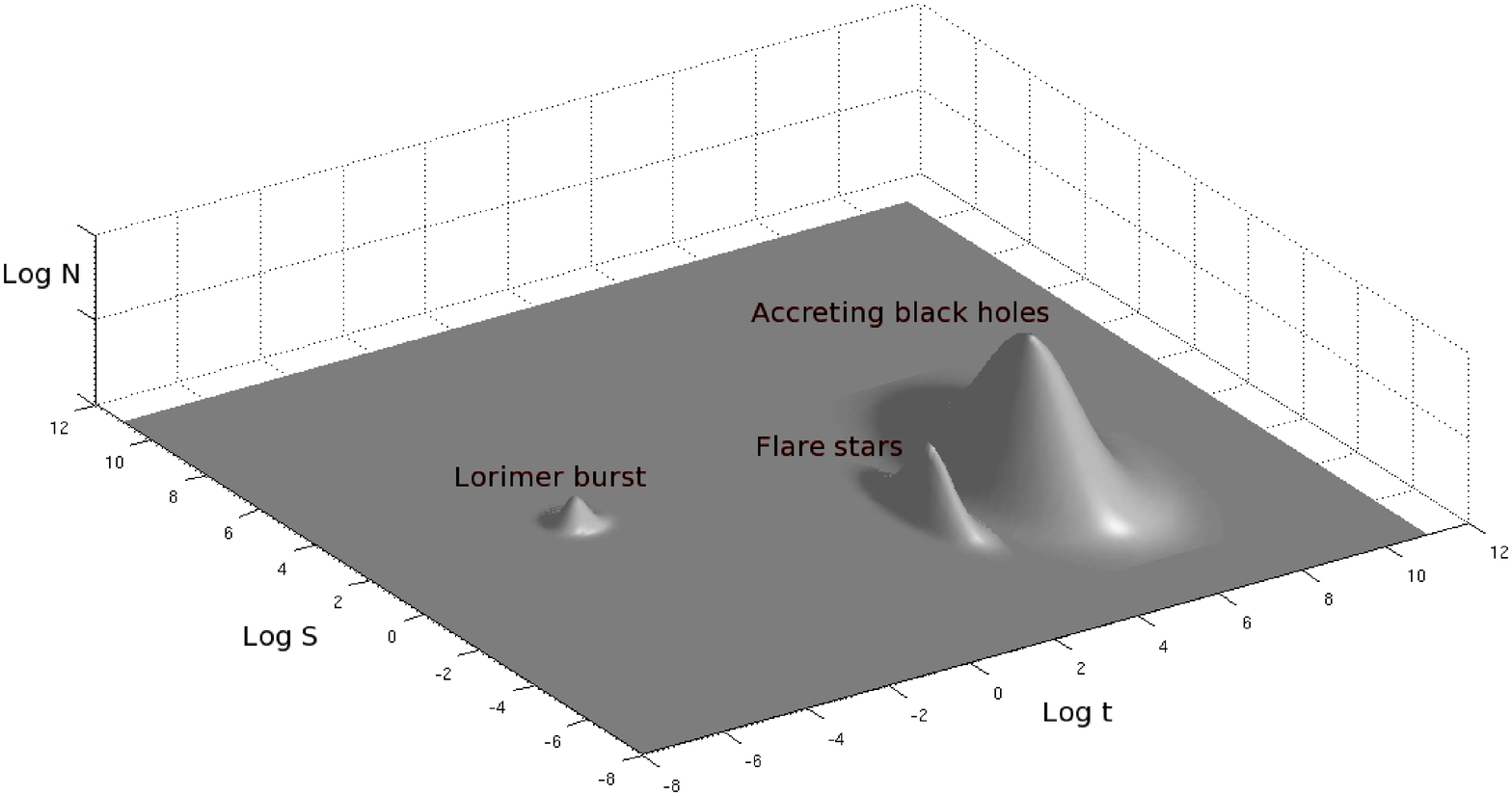}
\caption{A sketch of possible $\log N$ -- $\log S$ -- $\log t$ space,
  illustrating the milti-dimensional nature of the problem, and how
  current approaches tend to collapse together astrophysically diverse
  populations.}
\label{lognlogslogt}
\end{figure}

More work is needed to fully characterise the $\log N$ -- $\log S$. A
top priority should be cadence, the $\log N$ -- $\log S$ should evolve
to incorporate the characteristic timescale of transient behaviour
i.e. $\log N$ -- $\log S$ -- $\log t$ (see sketch in
Fig.~\ref{lognlogslogt}).  Furthermore, this parameter space
(including cadence) should be characterised at a number of different
radio frequencies (as noted above, the MHz and GHz transient
populations could be significantly different), and also in different
directions.  For example, in the direction of the Galactic plane,
variable radio sources may be more abundant. In a given extragalactic
direction, interstellar scintillation may be more prominent (due to
increased electron content), and will also potentially increase short
timescale variability at low frequencies (e.g. see Hunstead 1972). It
may also be interesting to compare (or even normalise) the transient
source $\log N$--$\log S$, with the known radio source $\log N$ --
$\log S$. For example, Windhorst et al. (1985) have shown that above
$\sim$ 1 mJy source populations are dominated by active galactic
nuclei, below this star-forming galaxies are thought to dominate (also
see Georgakakis et al. 1999 and Carilli et al. 2003).

What shall we do with these huge numbers of radio transients when we
find them ? At the time of writing the current goal is to achieve
rapid follow-up at optical wavelengths, since this remains the best
single tool to characterise the nature of such events. However, this
strategy may need to be revised if we find that the population is
dominated by sources with no optical counterparts (see e.g. discussion
in Bower et al. 2007, Ofek et al. 2010). 

\section{Conclusion: Route Inondable}

The benchmark study of Bower et al. (2007) is five years old and
remains the standard for our understanding of the transient sky at GHz
facilities. Several subsequent radio transient searches have delivered
upper limits or detections which are consistent with this, but neither
significantly refine the statistics nor detect sources in anything
close to real time, crucial for multiwavelength follow up and our
understanding of the astrophysics of these sources. The new generation
of radio facilities and approved key science projects should change
all of this, with near real time reporting of thousands of transient
events per year. This in turn will provide an extremely rich resource
for multiwavelength follow up, and the possible opening of a new
window on the high-energy Universe. Prepare for the flood.

\bgroup\small
\noindent{\em Note in manuscript:} Since this manuscript was written,
Frail et al. (2011) have reported that the Bower transient rates, on
which the estimates in this paper were largely based, may have been
overestimated by up to an order of magnitude. In this case the rates
predicted here may be similarly overestimated, but still correspond to
between thousands and millions per year, depending on the facility.
\egroup

\section*{Acknowledgements}

The authors would like to thank all of those with whom they have
discussed radio transients over the years, not least Ben Stappers,
Ralph Wijers and the rest of the LOFAR Transients Key Science Project,
Tara Murphy, Joe Lazio, Geoff Bower, Bryan Gaensler, Jean-Pierre
Macquart and Patrick Woudt and the rest of {\em ThunderKAT}. We would
also like to thank the editors for detailed comments which helped
improve this manuscript.


%

\label{lastpage}
\end{document}